\documentclass[aps,prb]{revtex4}   
\usepackage{graphicx}

\usepackage{amsmath}    
\usepackage{graphicx}

\begin{document}

\title{ \textit{Ab Initio} Construction of Interatomic Potentials for Uranium Dioxide Across all Interatomic Distances}
\author{P. Tiwary}
 \email{pt@caltech.edu}   
 \affiliation{Division of Engineering and Applied Sciences, California Institute of Technology, Pasadena, California 91125, USA}
\author{N. Gr{\o}nbech-Jensen}
 \email{ngjensen@ucdavis.edu}   
\affiliation{Department of Applied Science, University of California, Davis, California 95616, USA}
\author{A. van de Walle}
 \email{avdw@caltech.edu}  
\affiliation{Division of Engineering and Applied Sciences, California Institute of Technology, Pasadena, California 91125, USA}
\date{\today}

\begin{abstract}
We provide a methodology for generating interatomic potentials for use 
in classical molecular dynamics simulations of atomistic phenomena occurring at energy scales ranging 
from lattice vibrations to crystal defects to high energy collisions. A rigorous method to objectively 
determine the shape of an interatomic potential over all length scales is introduced by building upon a 
charged-ion generalization of the well-known Ziegler-Biersack-Littmark universal potential that provides the short- 
and long-range limiting behavior of the potential. At intermediate ranges the potential is smoothly adjusted by fitting to 
{\it ab initio} data. Our formalism provides a complete description
of the interatomic potentials that can be used at any energy scale, and thus, eliminates the inherent  ambiguity of splining different potentials generated to study different 
kinds of atomic materials behavior. We exemplify the method by developing rigid-ion potentials for 
uranium dioxide interactions under conditions ranging from thermodynamic equilibrium to very high atomic energy
collisions relevant for fission events.

\end{abstract}
\maketitle

\begin{figure}[htp]
\centering
 \includegraphics[width=86mm,height=55mm]{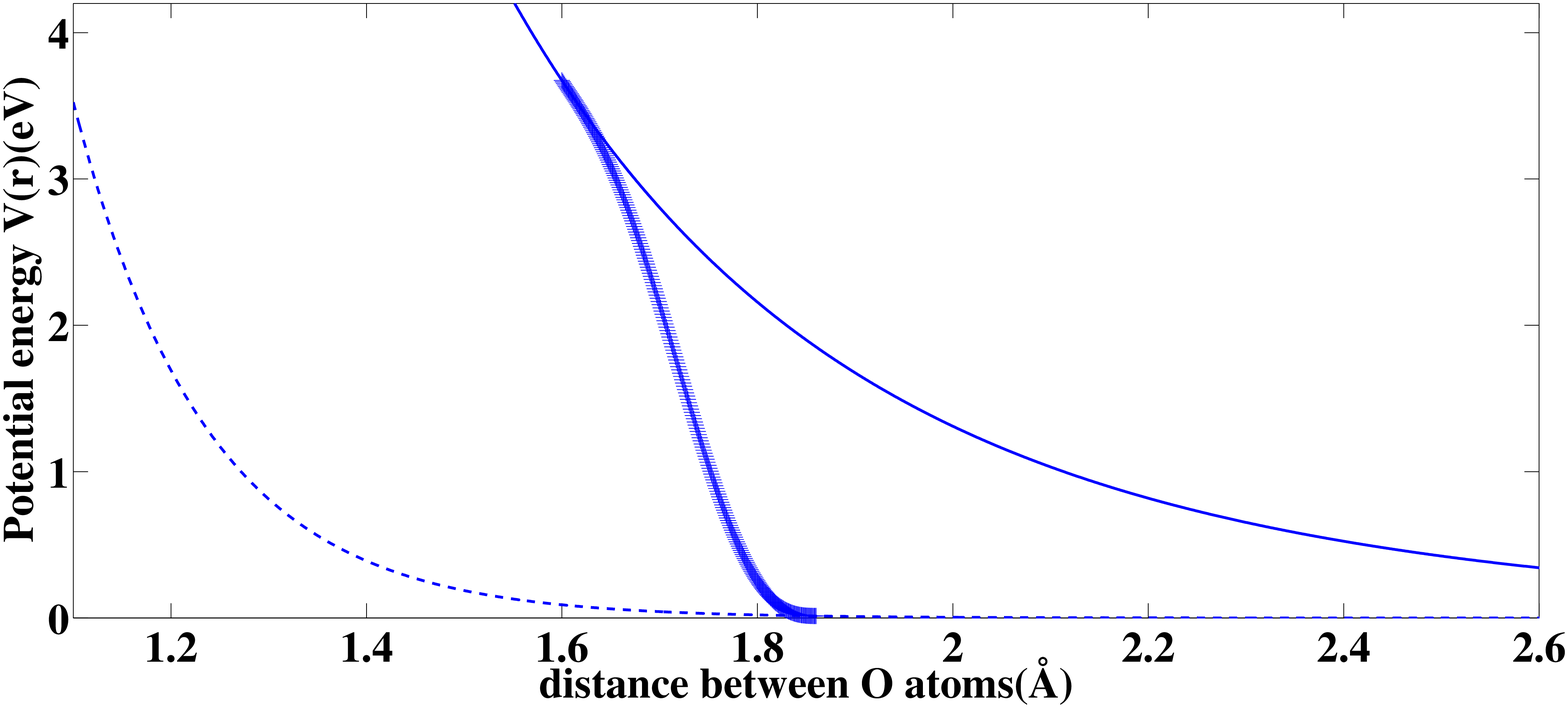}
 \includegraphics[width=86mm,height=55mm]{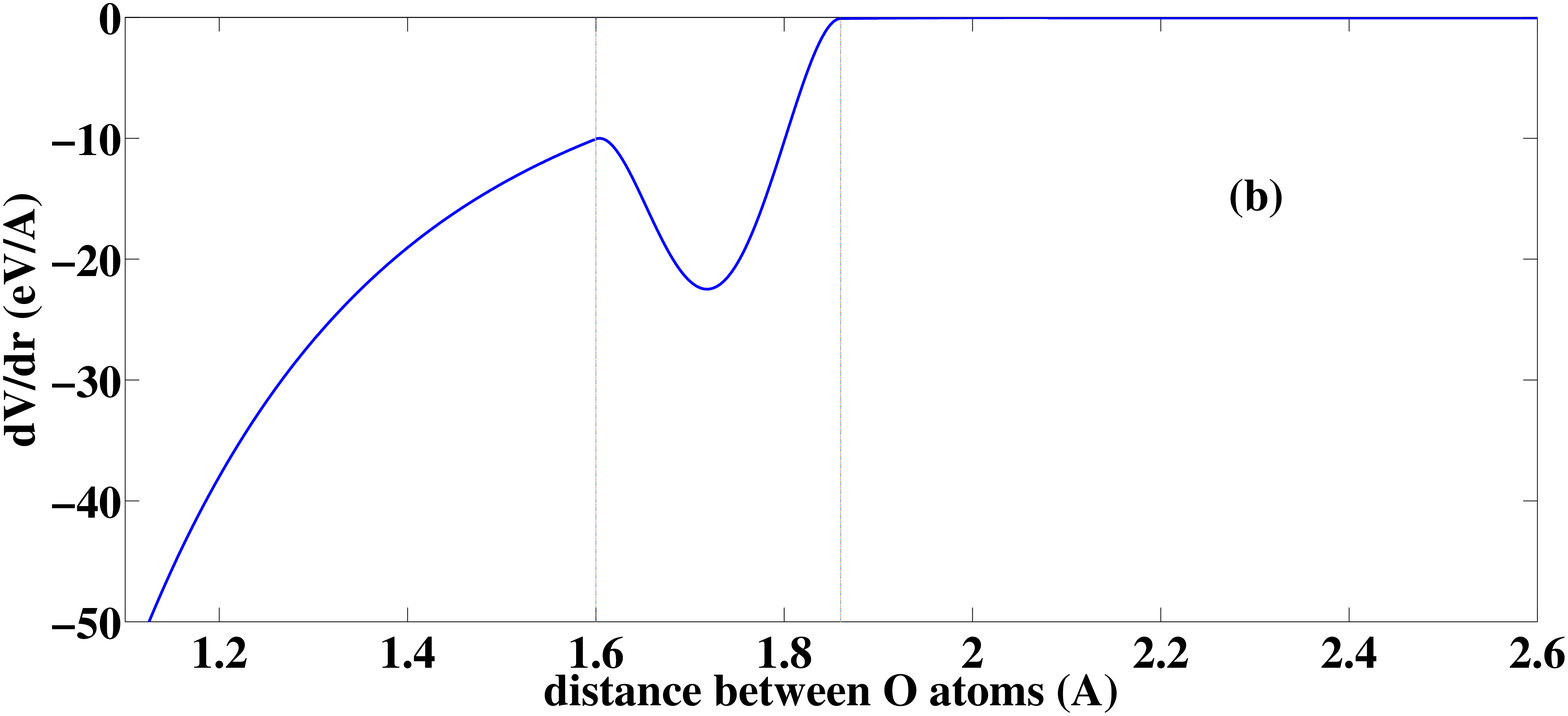}
 \includegraphics[width=86mm,height=55mm]{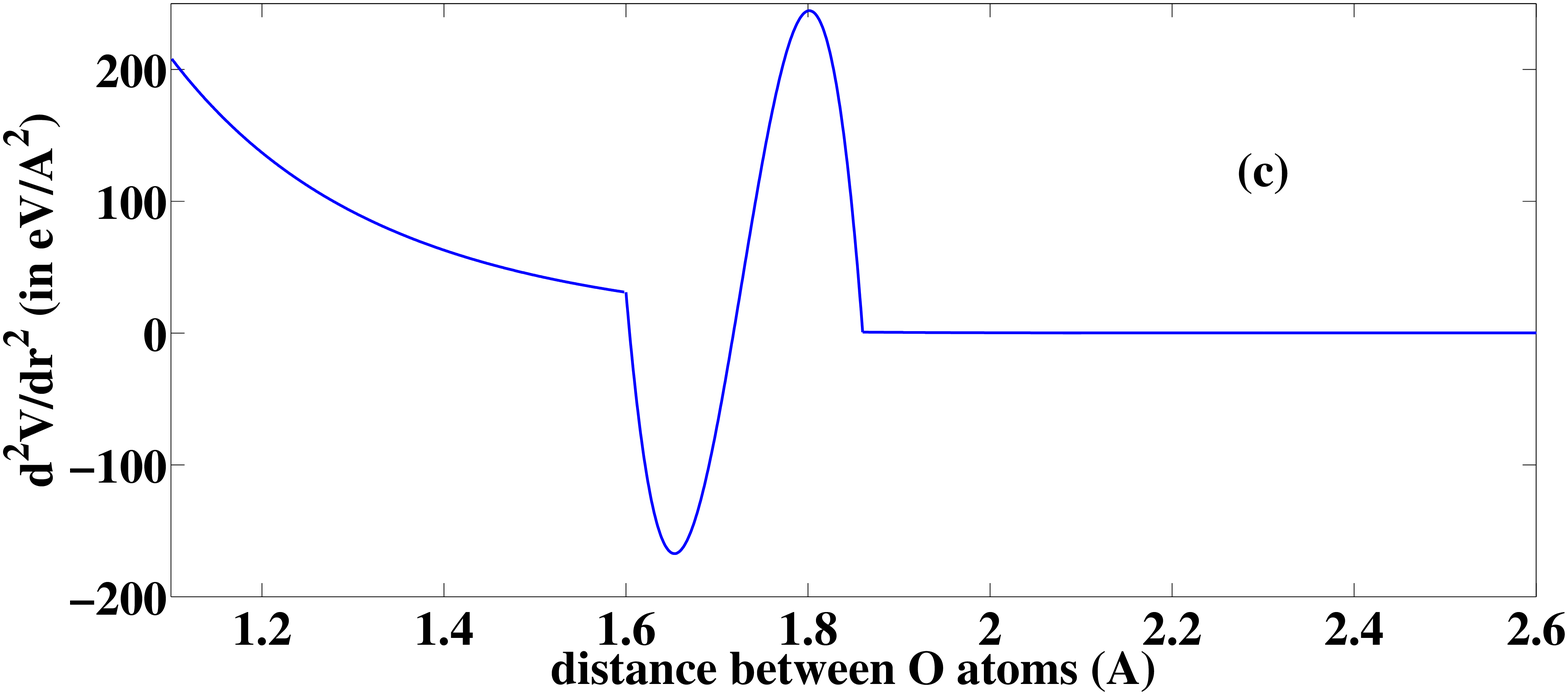}
  \caption[1]
{(a) ZBL screened potential (solid line) and Morelon \textit{et al.} potential (dashes), joined together by a 5th order polynomial (plus signs), for the case of two Oxygen atoms. (b) First derivative of the net potential resulting from the same. (c) Second derivative of the potential. Since the spline was not fit to any data, one can not decide whether the resulting behavior in (b) and (c) is correct or spurious.}
\label{fig:f1}
\end{figure}

Molecular Dynamics (MD) simulations provide a convenient tool for studying the dynamics of large atomic
ensembles, provided that the dynamics of interest is not of a duration that makes the simulation time
impractical. There are many examples of beautiful applications of how MD has been able to provide
detailed insight into the dynamics and statistics of materials behavior. One contemporary interest
is the field of high energy radiation damage of crystalline materials, such as nuclear fuel. A core material
of interest in this context is uranium dioxide, and one of the important aspects of the interest in this
material is to understand the evolution and statistics of atomic displacement cascades due to high
energy radiation \cite{Brutzel_rad_damage}. Classical Molecular Dynamics is ideally suited for this kind of
study since it strikes a fine balance between being coarse enough to simulate the spatial scale necessary to
represent the extent of a damage cascade due to, e.g., a 100 MeV atomic collision with being detailed enough
to retain the atomic structure of the material. The high energy range of
the potential (short range in atomic separation) is consistent with the well-accepted Ziegler-Biersack-Littmark (ZBL) universal pair potentials \cite{zbl}, which treat close range atomic interactions as screened Coulomb forces between the nuclei. However, the complexity of the true interatomic interactions
cannot be fully represented in an efficient manner by a simple classical functional form. Thus, one needs
to develop a set of essential interaction features that are necessary for a given application. This is a
particularly challenging exercise for radiation damage simulations due to the disparate scales of
energies involved. Therefore,
interatomic potentials, suitable for this purpose, are typically constructed by smoothly joining different types of interactions. At medium to long range distances, a traditional potential (e.g., Buckingham, electrostatic etc.) fitted to a variety of thermodynamic and structural data is used.
At short-ranges, accurate potentials are developed by fitting to \textit{ab initio} data. The ZBL universal potential is one such very popular pair-potential developed by Ziegler \textit{et al.} in the 1980's, as a generic function of the atomic numbers of the species involved.\cite{zbl}
Although each type of interaction is directly determined through fitting, the determination of a suitable
spline that smoothly joins these two pieces is a highly non-unique process.
Splining leads to an inherent ambiguity in the behavior of the complete potential, since the exact cutoff distances and the spline's algebraic form will have consequences for how large the ``cores'' of the atomic interactions are and how the potential behaves in the region of transition. This, in turn, will have a
significant impact on the ion trajectory and damage production one is ultimately interested in. Such
ambiguity is illustrated in Figure \ref{fig:f1}. Of course, a complete description of the dynamics of the system will also require the inclusion of a suitable electronic stopping model\cite{cai,brandt,zbl,barberan,wilson,nastasi}.

In the present communication we introduce a rigorous method to unambiguously determine the form of the potential over all distances using {\it ab initio} data. Although our approach is applicable to any material system, we illustrate it using the important example of the nuclear fuel UO$_2$, which has been the focus of several detailed computational investigations,
both through \textit{ab initio}\cite{geng2008} and MD\cite{brutzel_md,govers1,govers2} methods, due to its critical importance in the nuclear industry.

Our approach is to first generalize the universal ZBL potential to include charged ions that behaves correctly in both short range and long range limits. The advantage of building upon the ZBL formalism is that our potential automatically inherits the well-tested ability of the ZBL potential to describe high-energy scattering phenomena associated with the short-range behavior of the potential. This generalized potential smoothly interpolates between these two regimes over a physically-motivated length scale that is based on atomic orbital sizes instead of necessitating a user-specified transition radius for the electrostatic interactions at short ranges to prevent double-counting.
The only component that remains to be determined by fitting is then the medium-range energy contribution associated with chemical bonding. As this contribution is only significant over a relatively small range of distances it is possible to introduce lower and upper cutoffs, where this contribution must smoothly vanish.
Importantly, these physically motivated cutoffs can be fitted to {\it ab initio} data and are therefore no longer arbitrary; unlike in the current practice.

The ZBL potential\cite{zbl}, which properly accounts for the screening of nuclear charge by the electronic clouds as a function of interatomic distances, is built on considering two interacting spherically
symmetric rigid electron clouds.  In this spirit, we also consider two interacting spherically
symmetric rigid electron clouds with electron densities determined {\it ab initio} and fitted to a sum of Slater functions.\cite{slater} This approximation is valid beyond distances where electron clouds overlap and
chemical bonds form. For short distances (i.e., distances less than chemical bond lengths) we obtain the energy as a function of distance between any two atoms through first order perturbation
theory. It was demonstrated by Ziegler \textit{et al.} that more sophisticated self-consistent field calculations incorporating the distortion of electronic clouds did not lead to any significant differences in the resulting interatomic potential at short distances.\cite{zbl} Thus, since we employ the same electronic density and the same screening function (ratio of the actual atomic potential at some radius to the potential caused by an unscreened nucleus) as used by Ziegler \textit{et al.}, we recover the ZBL potential at these short distances, as we will show later. 

We consider two spherically symmetric charge densities per unit volume $\rho_1(r)$ and $\rho_2(r)$, with central point charges of $Z_1e$ and $Z_2e$ respectively, $\rho_1(r)$ and $\rho_2(r)$ being normalized to equal $(Z_1+q_1)e$ and $(Z_2+q_2)e$ respectively. We further think of the point charge $Z_1e$ as being made of two point charges, $(Z_1+q_1)e$ and  $(-q_1)e$; $Z_2e$ is similarly decomposed into $(Z_2+q_2)e$ and  $(-q_2)e$. $q_1e$ and $q_2e$ here denote the net ionic charges on atoms 1 and 2 respectively. We make this decomposition so that the Coulombic interaction term naturally arises in the expression for the net interaction potential $V(r)$,
	\begin{equation}
	 V(r) = ZBL_{Z_1+q_1,Z_2+q_2} (r) + {{q_1q_2e^2}\over{4\pi\epsilon_0r}} + t_1 + t_2 \; ,
	\end{equation}
where  $ZBL_{Z_1+q_1,Z_2+q_2} (r)$ denotes the ZBL form of interaction between two neutral atoms having atomic numbers ${Z_1+q_1}$ and ${Z_2+q_2}$, but using the screening length for $Z_1$ and $Z_2$. This is because the screening length is governed by the electronic charge distribution close to the nucleus and not far away from it.  Since the extra charges $q_1$ and $q_2$ have been added to the valence shells of neutral atoms with atomic numbers $Z_1$ and $Z_2$, we do not change the screening length. This was found critical in recovering the standard short-range ZBL potential. $t_1$  denotes the interaction between the point charge $-q_1e$ and the ${(Z_2+q_2)}e$ point charge plus electron cloud system of the atom 2, and is given by
\begin{equation}
	 t_1(r) = {{-q_1e^2}\over{4\pi\epsilon_0}} {[{{Z_2+q_2}\over r}-{1\over r}{\int_0^{r}4\pi s^2 \rho_2(s) ds}-{\int_r^{\infty}4\pi s\rho_2(s) ds}]} \; ,
	\end{equation}
while $t_2$ is the converse remaining point-atom interaction, expressed similarly.
	Eq.(1) is  correct for atomic separations smaller than the 'chemical bond length' (where it recovers the original neutral atom ZBL) and for very large atomic separations as well. In the latter case, only the 2nd term in Eq.(1) survives as we show below. 

\begin{figure}[htp]
\centering
 \includegraphics[width=86mm, height=55mm]{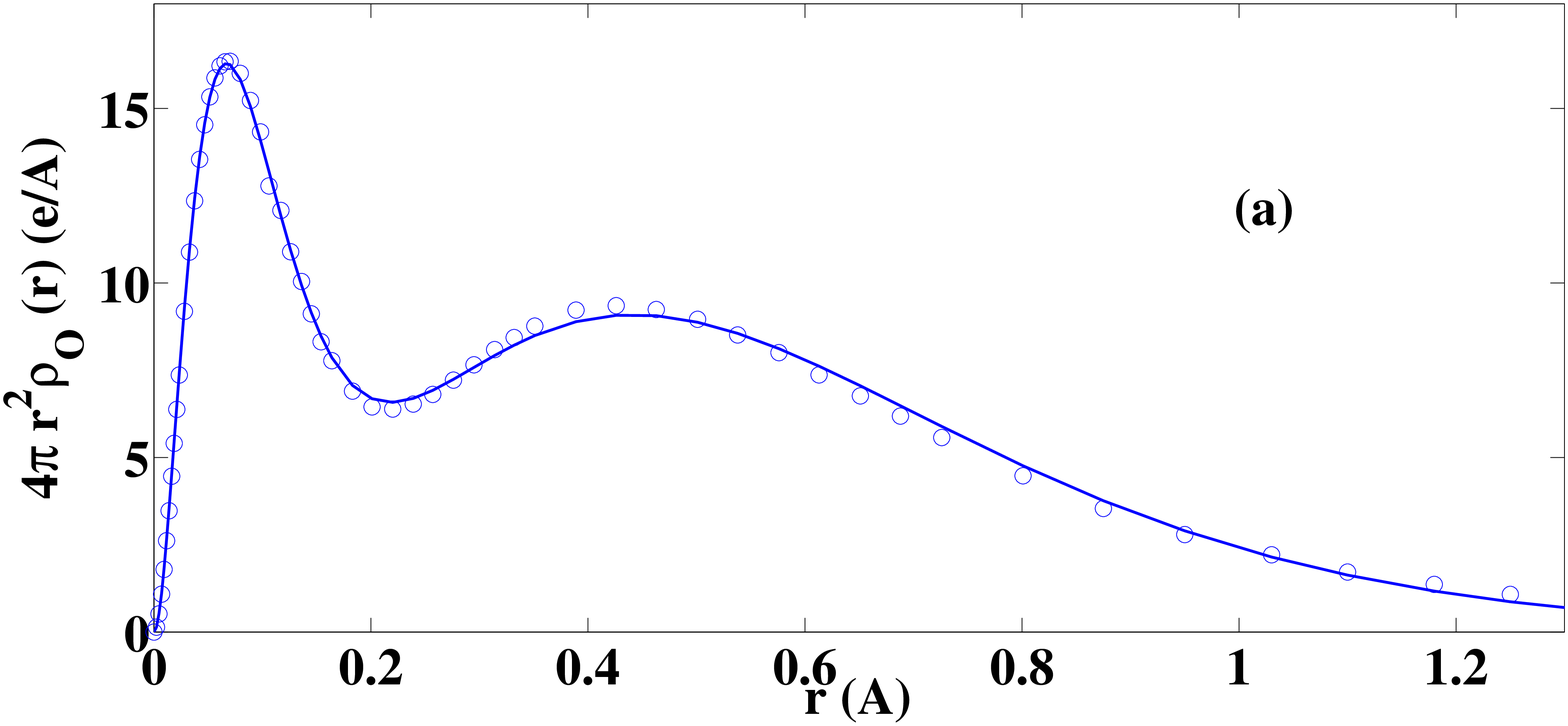}
 \includegraphics[width=86mm, height=55mm]{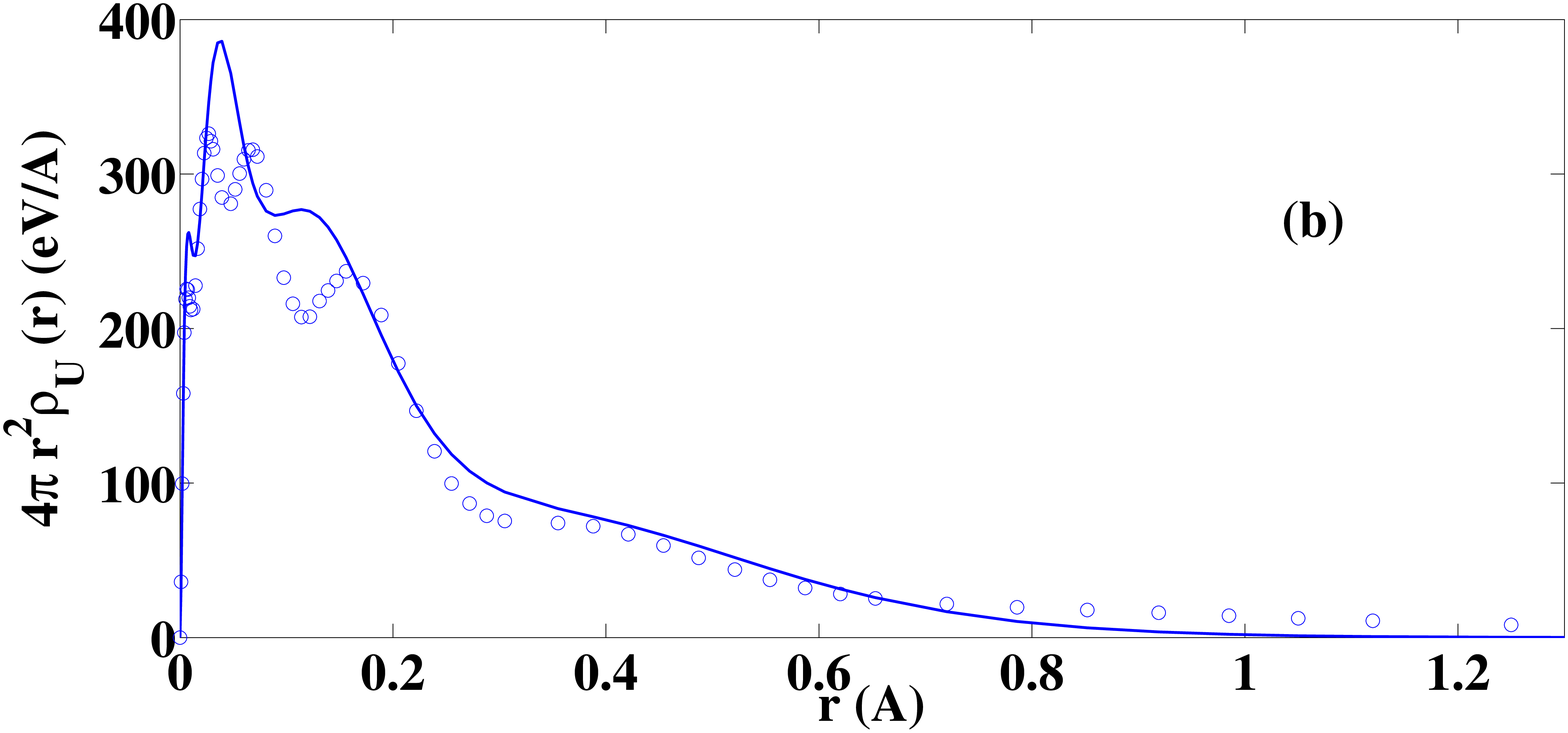}
  \caption[2]
{Charge density times $4\pi r^2$ fitted to sum of Slater functions for (a) Oxygen and (b) Uranium. Open circles denote values used by Ziegler \textit{et al.},\cite{zbl} while the solid lines indicate our fit using sum of Slater functions. The resultant error in the short range interatomic potential as compared to ZBL's original potential was well within the latter's standard deviation for both (a) and (b). Thus trying to capture more peaks for Uranium, by introducing more Slater functions, was not necessary. The coefficients used here are provided in Table \ref{table:slater}. }
\label{fig:f2}
\end{figure}

	The task now is the determination of $\rho_O(r)$ and $\rho_U(r)$. It is here important to notice that
the potential for very small interatomic separations is only as good as the ZBL form (see Eq.~(1)). Thus, we use the charge densities employed for ZBL, which are primarily Hartree-Fock-Slater atomic distributions for most of the atomic pairs. We fit the numerical data for charge density used by Ziegler \textit{et al.} to a sum of Slater functions.
While the density $\rho(r)$ is known to be a monotonic decreasing function of radial distance for all atoms, the graph of  $4\pi r^2\rho(r)$ exhibits a number of peaks (see Figure \ref{fig:f2}) corresponding to atomic shells. To ensure the best possible accuracy, we fit to  $4\pi r^2\rho(r)$ because this is the quantity entering Eq.(1).
We find that 2 and 4 Slater functions are sufficient to capture the behavior of $4\pi r^2\rho(r)$ for neutral Oxygen and Uranium atoms (see Figure \ref{fig:f2}), i.e.,
	\begin{equation}
	 \rho_O(r) = a_1e^{-k_1r}+a_2re^{-k_2r}
	\end{equation}

\begin{equation}
	 \rho_U(r) = b_1e^{-l_1r}+b_2re^{-l_2r}+b_3r^2e^{-l_3r}+b_4r^3e^{-l_4r} \; .
	\end{equation}

Fitting the ZBL charge density with Slater functions does not exactly ensure that the areas under the $4\pi r^2 \rho(r)$ curves for Oxygen and Uranium are 8 and 92 respectively.
The Slater function fits therefore needed a slight modification in their pre-factors. In addition, the pre-factors in the above two equations as reported in Table \ref{table:slater} have been multiplied by 10/8 for Oxygen and 88/92 for Uranium because we are interested in the electronic cloud of the ionic species and not of the neutral atoms. We also experimented with more sophisticated corrections (e.g., using noble gas densities instead) but this did not change the results by more than the intrinsic accuracy of the ZBL potential.
By using Eqs. (3) and (4) in Eq.(1) and performing the integrations, we obtain the following pair potentials for Oxygen-Oxygen, Uranium-Uranium and Oxygen-Uranium respectively:

\begin{equation}
	 V_{OO}(r) = ZBL_{10,10} (r) + {{(-2)(-2)e^2}\over{4\pi\epsilon_0r}} - {{4e^2}\over{4\pi\epsilon_0}}[{10\over r}-{4\pi \over {e}} {f_{OO}(r)}]
	\end{equation}

																\begin{equation}
	 V_{UU}(r) = ZBL_{88,88} (r) + {{(4)(4)e^2}\over{4\pi\epsilon_0r}} + {{8e^2}\over{4\pi\epsilon_0}}[{88\over r}-{4\pi \over {e}} f_{UU}(r)]
	\end{equation}
	
	\begin{equation}
	 V_{OU}(r) = ZBL_{88,10} (r) + {{(4)(-2)e^2}\over{4\pi\epsilon_0r}} - {{2e^2}\over{4\pi\epsilon_0}}[{88\over r}-{4\pi \over {e}} f_{UU}(r)] +{{4e^2}\over{4\pi\epsilon_0}}[{10\over r}- {4\pi \over {e}} f_{OO}(r)]
	\end{equation}
	
	where we have
	
	\begin{equation}
	 f_{OO}(r) =   {6a_2\over rk_2^4}-  {a_2e^{-k_2r}\over {rk_2^4}}[6+4k_2r+k_2^2r^2] + {a_1\over {rk_1^3}}[2-2e^{-k_1r}-k_1re^{-k_1r}] 
	\end{equation}

\begin{multline}\label{mult}
 f_{UU}(r) =  {120b_4\over rl_4^6}-  {b_4e^{-l_4r}\over {rl_4^6}}[120+96l_4r+36l_4^2r^2+8l_4^3r^3+l_4^4r^4]+\\
 {24b_3\over rl_3^5}-  {b_3e^{-l_3r}\over {rl_3^5}}[24+18l_3r+6l_3^2r^2+l_3^3r^3]+\\
 {6b_2\over rl_2^4}-{b_2e^{-l_2r}\over {rl_2^4}}[6+4l_2r+l_2^2r^2] + 
  {b_1\over {rl_1^3}}[2-2e^{-l_1r}-l_1re^{-l_1r}]   
\end{multline}

\begin{figure}[htp]
\centering
 \includegraphics[width=86mm, height=55mm]{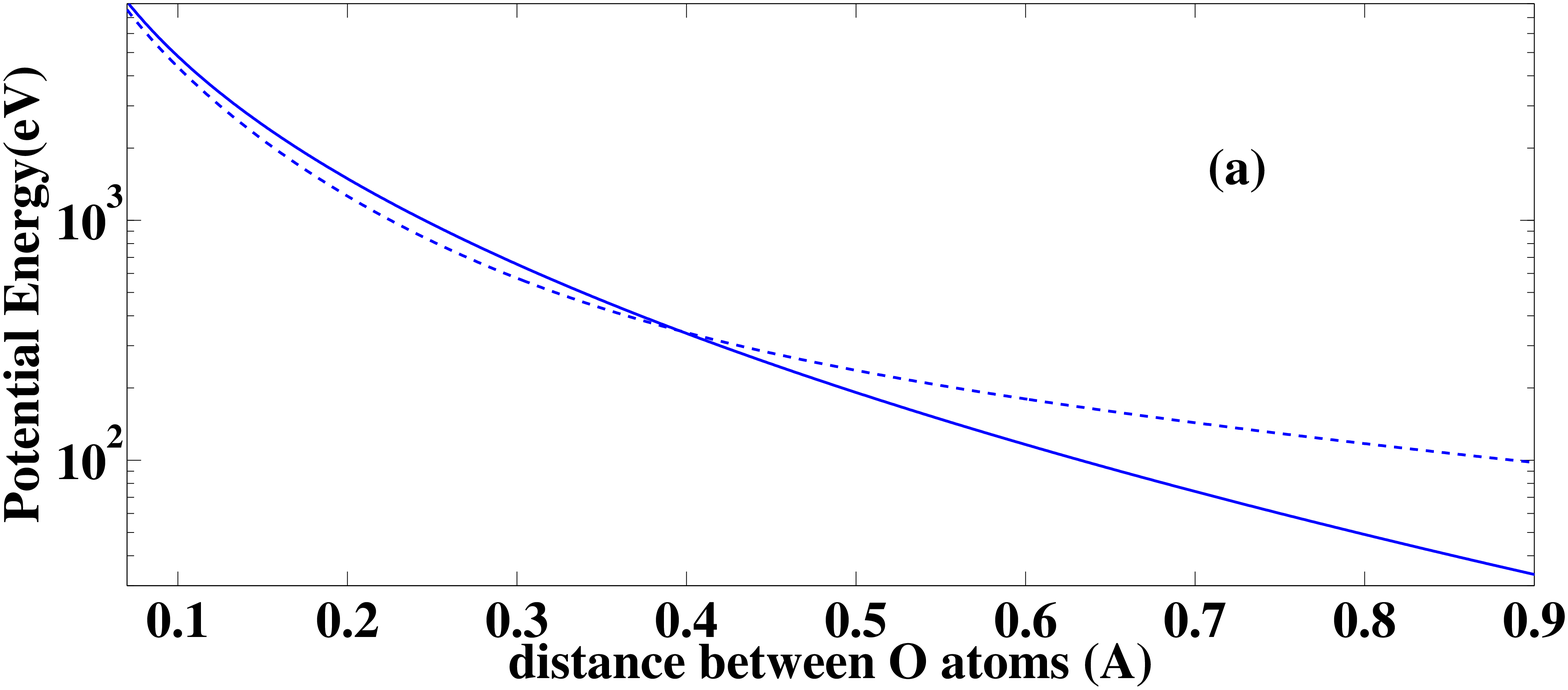}
 \includegraphics[width=86mm, height=55mm]{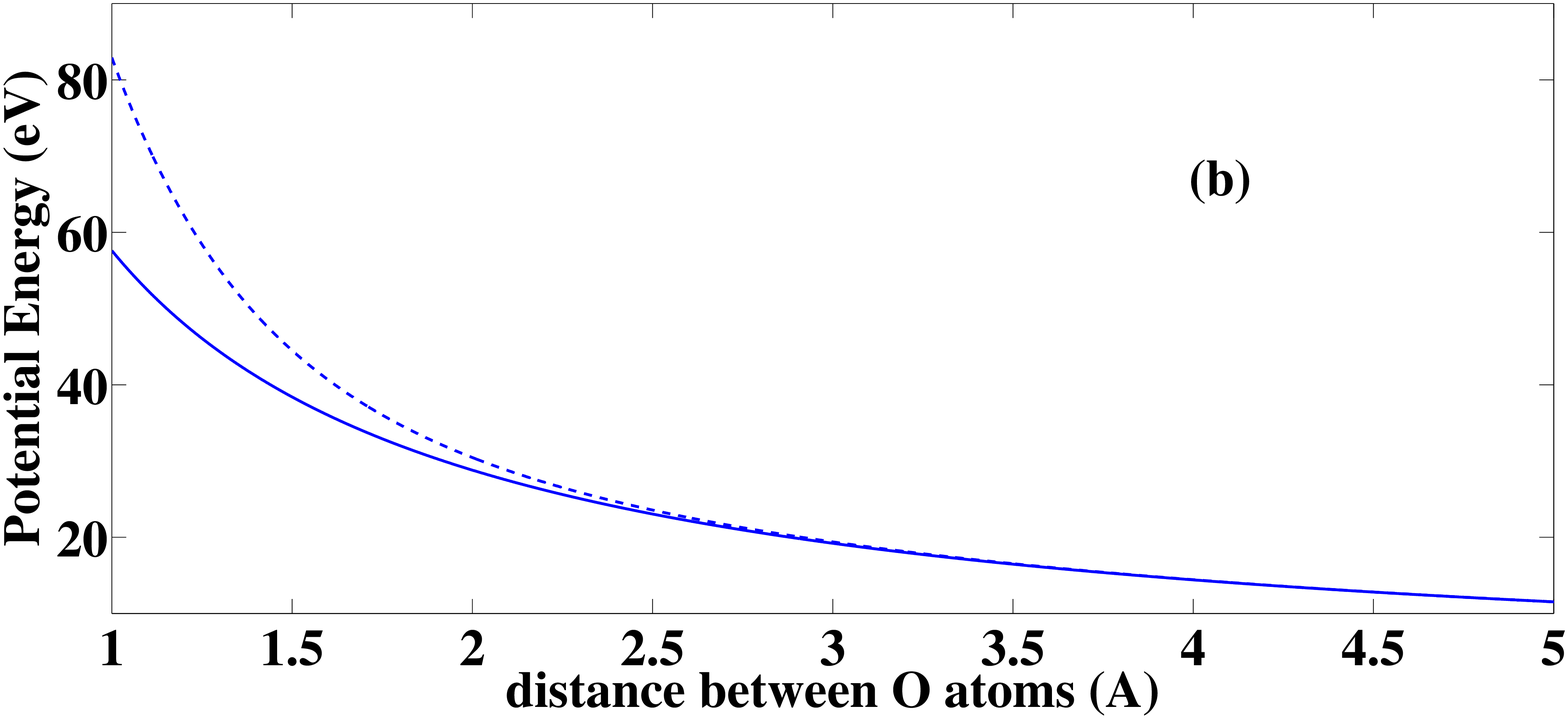}
  \caption[3]
{Comparison of our analytical potential form (dashed line) with (a) the currently used neutral atom ZBL interaction (solid line) for small distances, and (b) the ionic coulombic interaction (solid line) between two (-2$e$)point charges for large distances for the case of two Oxygen atoms.
}
\label{fig:f3}
\end{figure}

We illustrate in Figure \ref{fig:f3} how closely the potentials given in Eqs. (5) - (7) match, for small $r$, the neutral atom ZBL, and for large $r$, the relevant Coulombic interaction.

As for any empirical potential, there is an intermediate distance range for which the interactions follow neither
a ZBL nor a purely Coulombic form. A correction term is thus needed for this regime. We find this correction term by fitting to an extensive database of GGA+\textit{U ab initio} calculations on UO$_2$.
GGA+$U$ is known to provide electronic and magnetic behaviors of UO$_2$\ that are consistent with experiments \cite{laskowski}, and a correct treatment of the localized and strongly correlated 5$f$ electrons of Uranium.\cite{geng2007} Our \textit{ab initio} calculations also take into account the experimentally observed noncollinear antiferromagnetic magnetic moment ordering and the Oxygen cage distortion in UO$_2$.\cite{ikushima} Therefore, in addition to capturing correct elastic and defect properties, our potential also covers a much more vast energy landscape in the material due to the richness of the \textit{ab initio} data used. We fit to a database obtained by GGA+$U$ calculations with the projector-augmented-wave method implemented in the VASP\cite{VASP} package. In the GGA+$U$ approximation, the spin-polarized GGA potential is supplemented by a Hubbard-like term to account for the strongly correlated 5$f$ orbitals.\cite{lda+u} We use the rotationally invariant approach to GGA+$U$ due to Dudarev \textit{et al.}\cite{dudarev1}, wherein the parameter \textit{U-J} is set to 3.99 eV. This is the generally accepted value for this parameter to reproduce the correct band structure for UO$_2$.\cite{dudarev2} The magnetic moments were allowed to be fully non-collinear. The \textit{ab initio} database so obtained comprises:

(i) isochoric relaxed runs on a 12 atom unit cell which was isometrically contracted and expanded by various amounts (i.e., equation of state calculations wherein each data point was calculated under constraint of constant cell volume), and for which an energy cutoff of 500 eV and a 8$\times$8$\times$8 k-point grid were taken; k-point convergence was ascertained before choosing this value for the k-point grid. The cell was allowed to relax in shape but not in size. Ionic relaxations were carried out until residual forces less than 0.01 eV/\AA were achieved. 

(ii) static (i.e., no ionic relaxation) runs on 96 atom 2$\times$2$\times$2 supercell in which one atom at a time (i.e., Oxygen or Uranium) was perturbed from its equilibrium position by varying distances in different directions. Energy cutoff was 500 eV. After performing convergence studies on the k-point grid, gamma point only version of VASP was found to be satisfactorily accurate for this. Note that any interactions between atoms and their periodic images do not systematically bias the fit of the potential because the same supercell geometry is used in both the {\it ab initio} and the empirical potential energy calculations.

(iii) 96 atom 2$\times$2$\times$2 supercell for the formation energies of three kinds of stoichiometric defects, namely Oxygen Frenkel pair, Uranium Frenkel pair and Schottky trio. The vacancies and the interstitials were taken as far from each other as the supercell would allow. The details of the calculations are the same as that for case (ii) above.  Correct prediction of defect energies has been given great importance in generating interatomic potentials for cascade simulations.

(iv) first-order transition states in a 2$\times$2$\times$2 supercell for the migration energy of Oxygen and Uranium vacancy and interstitial. Nudged Elastic Band(NEB) method\cite{neb} in conjunction with the climbing image method\cite{climbingimage} for determination of saddle point energy, as implemented in VASP, was used for this. 

With the \textit{ab initio} database so generated, we now fit the final potential forms as follows. 

\begin{equation}
	 V_{UU}(r) = ZBL_{88,88} (r) + {{(4)(4)e^2}\over{4\pi\epsilon_0r}} + {{8e^2}\over{4\pi\epsilon_0}}[{88\over r}-{4\pi \over {e}} f_{UU}(r)]
	\end{equation}

\begin{eqnarray}
	V_{OO}(r) &= &{{(-2)(-2)e^2}\over{4\pi\epsilon_0r}} +\left\{ 
\begin{array}{l l}
  ZBL_{10,10} (r) - {{4e^2}\over{4\pi\epsilon_0}}[{\frac{10}{r}}- {4\pi \over {e}} f_{OO}(r)] & \quad {0 < r \leq 1.17 \text{\AA}}\\
    5^{th}\; order \;polynomial & \quad {1.17 \text{\AA} < r \leq 2.28 \text{\AA} }\\
      3^{rd}\; order\; polynomial & \quad {2.28 \text{\AA} < r \leq 2.84 \text{\AA} }\\
  {-603.268\text{eV\AA}^6/r^6} & \quad {r > 2.84 \text{\AA} }\\ \end{array} \right. \nonumber
  \end{eqnarray}
\begin{eqnarray}
	V_{OU}(r) &=& {{(-2)(4)e^2}\over{4\pi\epsilon_0r}} \nonumber \\
	& + & \left\{ 
\begin{array}{l l}
  ZBL_{88,10} (r) + {{4e^2}\over{4\pi\epsilon_0}}[{\frac{10}{r}}- {4\pi \over {e}} f_{OO}(r)]  - {{2e^2}\over{4\pi\epsilon_0}}[{\frac{88}{r}}- {4\pi \over {e}} f_{UU}(r)] & \quad {0 < r \leq 1.42 \text{\AA}}\\
    5^{th}\; order \;polynomial & \quad {1.42 \text{\AA} < r \leq 1.70 \text{\AA}}\\
  394.391\text{eV}\exp(-r/0.534\text{\AA})-{1.5\text{eV}\text{\AA}^6/r^6} & \quad {r > 1.70 \text{\AA}}\\ \end{array} \right. \nonumber 
  \end{eqnarray}

\begin{figure}[htp]
\centering
 \includegraphics[width=86mm, height=55mm]{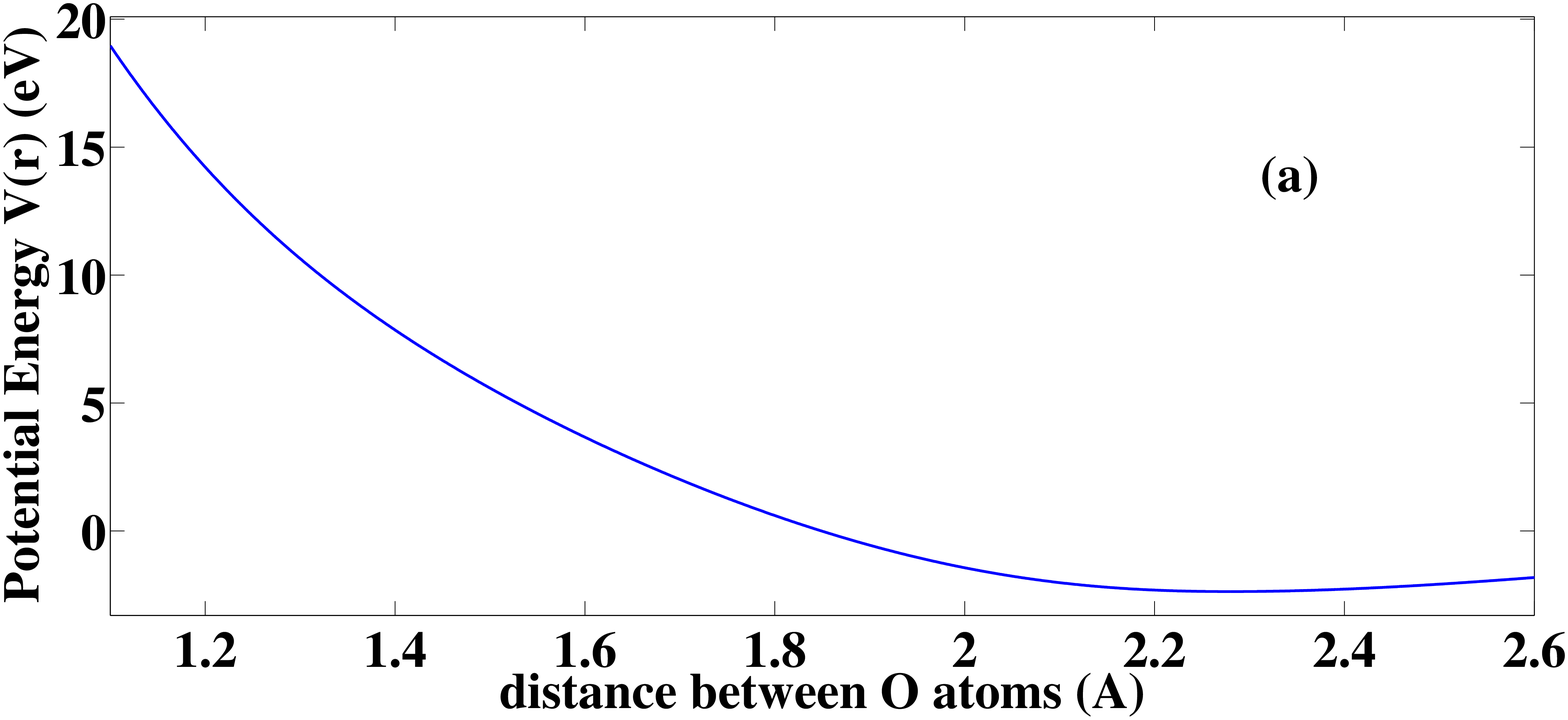}
 \includegraphics[width=86mm, height=55mm]{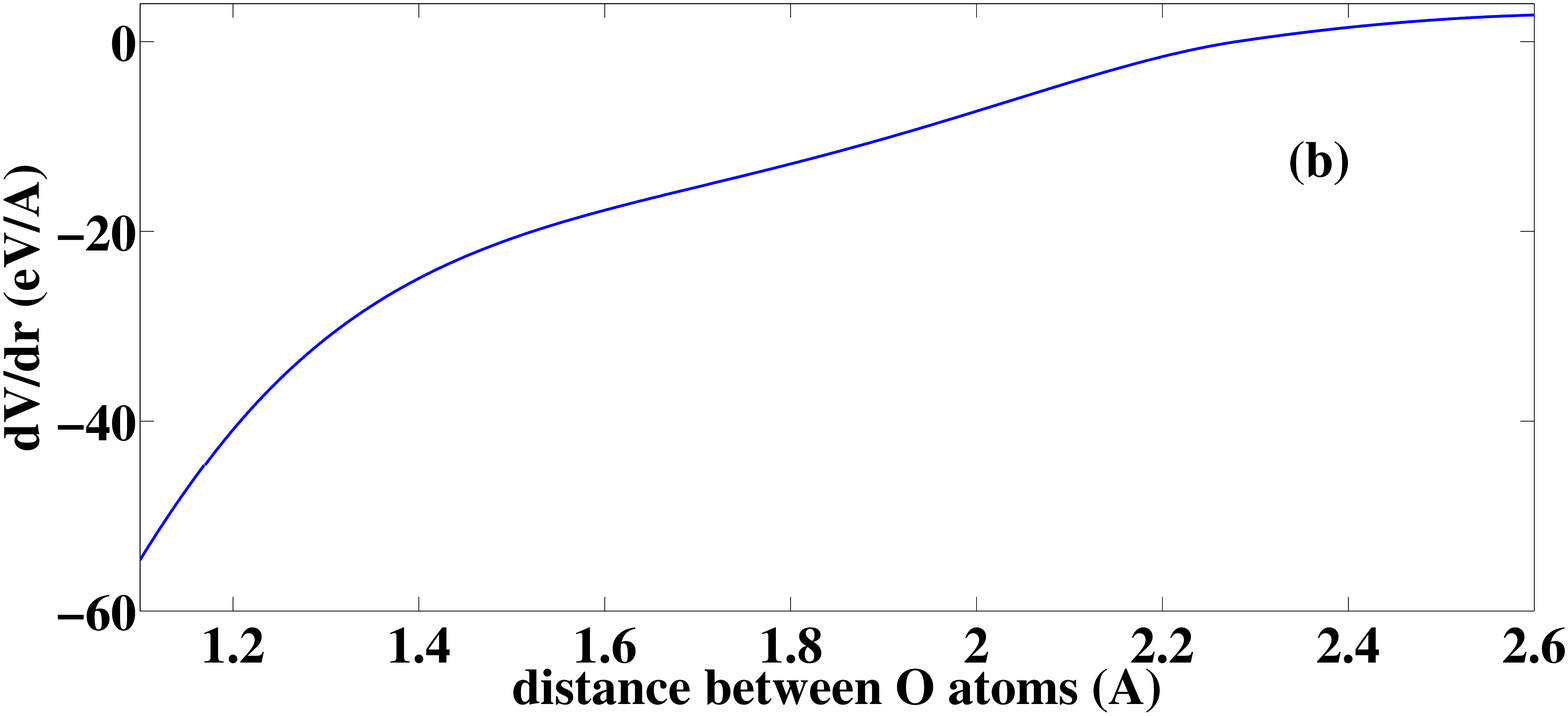}
 \includegraphics[width=86mm, height=55mm]{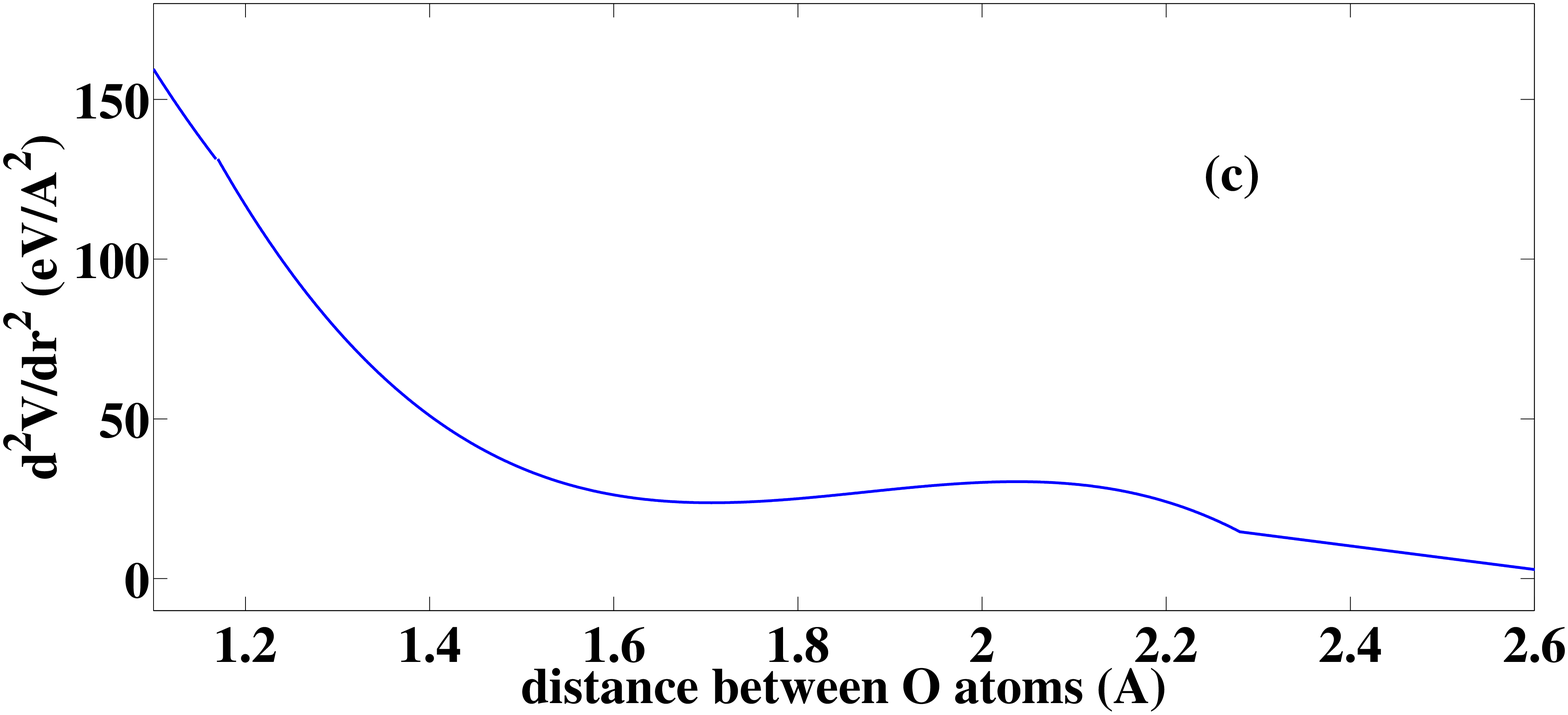}
  \caption[4]
{(a) The interatomic potential for Oxygen-Oxygen as per current work (Eq. 10). (b) First derivative of the net potential resulting from the same. (c) Second derivative of the potential. These are to be compared with Figure 1.}
\label{fig:f7}
\end{figure}

The long-range Coulomb terms in the interaction here were calculated through the standard Ewald summation technique. The upper cut-offs for all terms except the Coulombic in Eq. (10) may be chosen as per the availability of computational resources.
As seen from Eq. (10) we now have absolutely no splines for the U$^{+4}$-U$^{+4}$ interaction, reflecting the fact that
no chemical bonding takes place. There are splines in the other two interactions but these are now unambiguously determined 
since the respective cut-offs are not imposed but instead determined through fitting. The splines maintain continuity through the second derivatives of the potential and the specific form of the splines can be uniquely recovered from these conditions. Since they have been fitted to accurate \textit {ab initio} data, these splines do not introduce any spurious wriggles in the potential. For the interaction between two Oxygen ions, the potential has one (and only one) minimum at $r_{min} = 2.28$\AA, as may be seen from Figure \ref{fig:f7}. We have thus minimized the unphysical features of all of the interatomic potentials in UO$_2$, which were demonstrated for the particular case of Oxygen-Oxygen interaction in Figure \ref{fig:f1}. 
 
\begin{figure}[htp]
\centering
 \includegraphics[width=86mm, height=55mm]{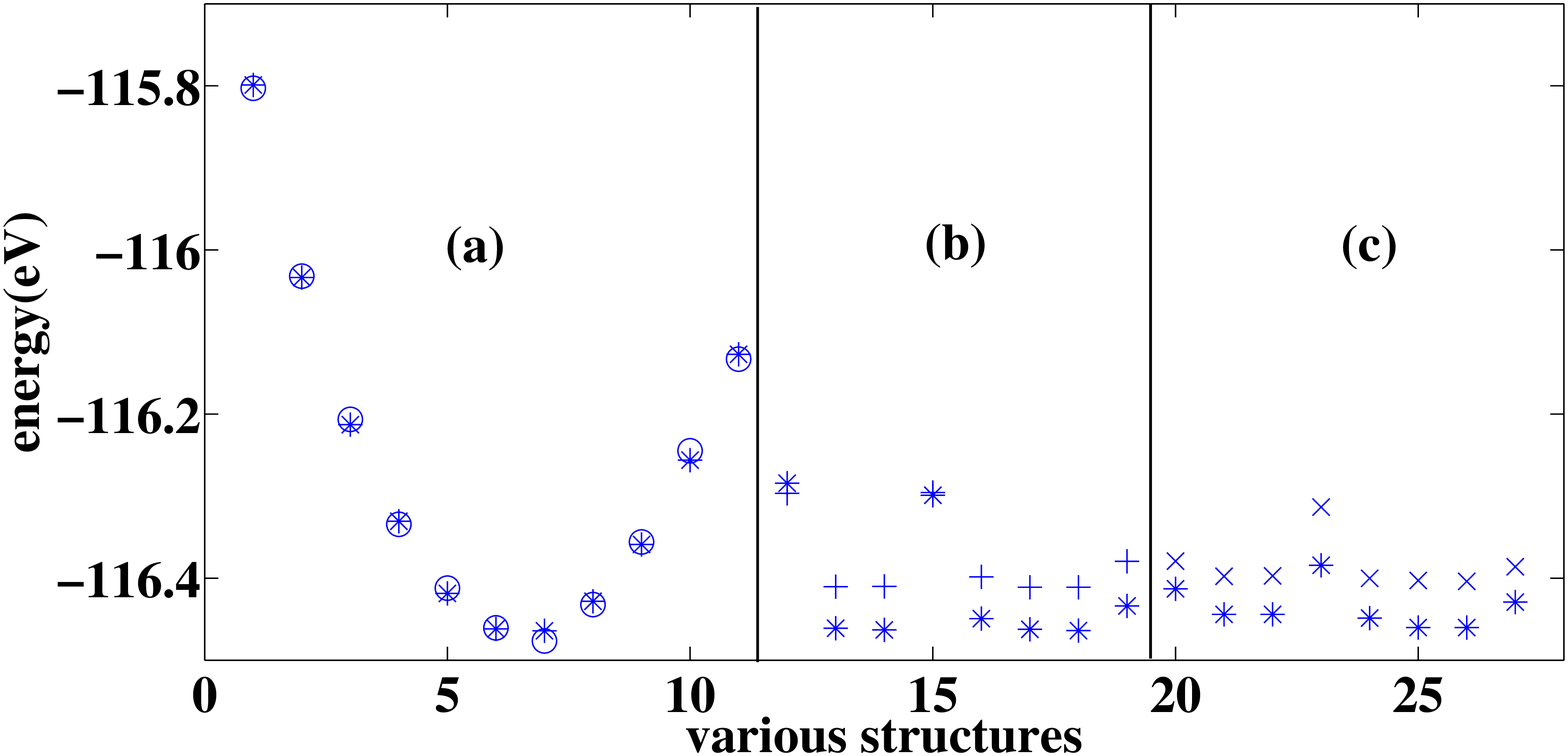}
  \caption[5]
{Quality of fit from our fitted potential for various \textit{ab initio} energies: (a) expansion/contraction (open circles), (b) Oxygen atom perturbation (plus signs), (c) Uranium atom perturbation (cross signs). Asterisks denote experimental data. For each of Oxygen and Uranium, the first four perturbations are along $<100>$ direction while the second four are along $<110>$ direction.
}
\label{fig:f4}
\end{figure}

  The downhill simplex method of Nelder-Mead was used to carry out the fitting.\cite{amoeba} The fitting involved minimizing the sum of the squares of the differences between the \textit {ab initio} energies and the energies predicted by the potential for all the classes of data points as detailed above. The package GULP\cite{GULP} was used for energy calculations and for atomic positions optimization. The quality of fit for the equation of state data and the perturbed atom data can be seen in Figure \ref{fig:f4}. The \textit{ab initio}/experimental and predicted defect formation/migration energies are compared in Table \ref{table:energies}, while Table \ref{table:elastic} lists the predicted ground state lattice parameter and other elastic properties as compared with the corresponding experimental values\cite{govers1,fink} (extrapolated accordingly) and with values obtained with the Morelon \textit{et al.} potential\cite{morelon_pot}. The agreement is very satisfactory.
 
\begin{figure}[htp]
\centering
\includegraphics[width=86mm, height=55mm]{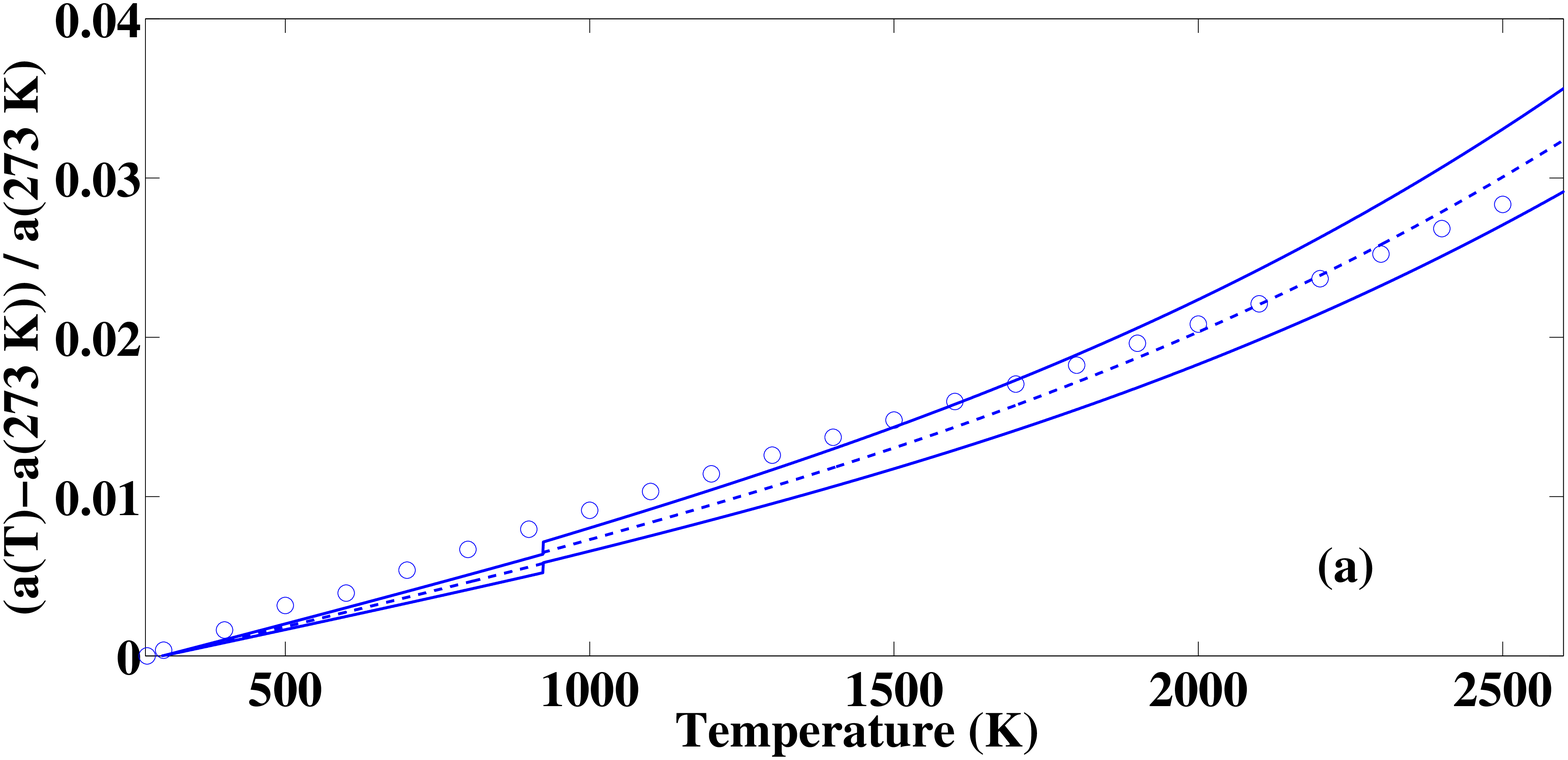}
\includegraphics[width=86mm, height=55mm]{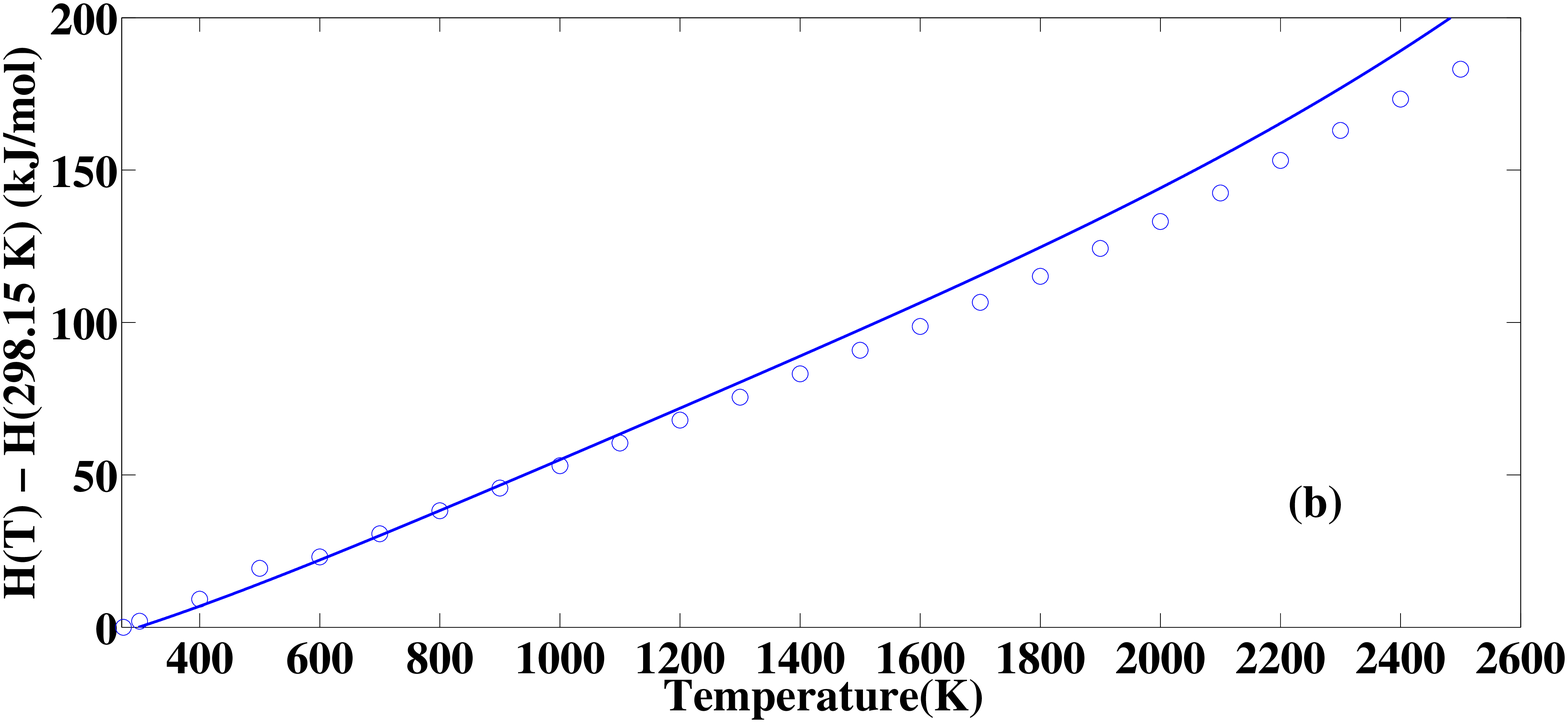}
  \caption[6]
{(a) Relative lattice parameter variation using potential from current work (open circles) compared with corresponding experimental values (dashed line).\cite{fink} The scatter in the experimental values is also shown (solid lines). (b) Enthalpy variation using potential from current work (open circles) compared with corresponding experimental values (solid line).\cite{fink} The scatter in the experimental values was less than 1 percent and is thus not shown.
}
\label{fig:f5}
\end{figure}

As a final validation of the developed potential we considered various dynamic properties by performing MD simulations in a constant number, pressure and temperature (NPT) ensemble comprising 6$\times$6$\times$6 unit cells. The system was equilibrated for 10.0 ps, while production runs were carried out for 100.0 ps, with a time step of 0.001 ps and sampling every 0.05 ps. The fluorite structure remained stable during all the runs we performed, up to temperatures of 2500 K. The properties we considered are the variation in the lattice parameter and the enthalpy as functions of the temperature. These are compared in Figure \ref{fig:f5} with the corresponding experimental data.\cite{fink} The quality is similar to what is given by the previous potentials\cite{morelon_pot, old_pot_1, old_pot_2}, as tabulated in the work by Morelon \textit{et al.}\cite{morelon_pot}

To summarize, we have shown a methodology for developing an interatomic pair potential such that it is appropriate for all relevant interatomic separations, without the need for any ambiguous splines. Splining
between regions of different characteristics is not just an inconvenience in the implementation of a potential in MD simulations, but also introduces an uncertainty regarding which distances, and by which functions, one realizes the splines. The potential we have focused on in this presentation, the nuclear fuel UO$_2$,
is just one example of a model material in which very relevant materials physics depends on accurate and reliable interactions over many orders of magnitude, and we have obtained the first complete description
that allows for direct simulations of damage cascades due to high energy radiation effects. The potential
has been generated based on a slight revision of the ZBL universal potential to account for ionic materials
with the intermediate interatomic distances fitted to a  broad database of \textit{ab initio} structural energies.
In view of these qualities, we expect it to be a very reliable potential for studying displacement cascades in
UO$_2$.

The authors would like to thank Dr. Byoungseon Jeon for pointing out typographical errors in the manuscript and for helpful suggestions. This research was supported by the US National Science Foundation through TeraGrid resources provided by NCSA under grant DMR050013N, through the U.S. Department of Energy, National Energy Research Initiative Consortium (NERI-C), grant DE-FG07-07ID14893, and through the Materials Design Institute, Los Alamos National Laboratory contract 25110-001-05.

\newpage

\begin{table}[htp]
\caption{ Values of coefficients in Slater functions in Eqs. (3) and (4)}
\centering
\begin{tabular}{c c c}
\hline\hline

\hline
       a$_1$ =        2799.625 e\AA$^{-3}$&
       a$_2$ =       211.038 e\AA$^{-4}$&
       k$_1$ =       30.76 \AA$^{-1}$ \\
       
       k$_2$ =       6.77 \AA$^{-1}$ &
        b$_1$ =   3092188.94 e\AA$^{-3}$&
       b$_2$ =   13255095.09 e\AA$^{-4}$ \\
       
       b$_3$ =   4982192.00 e\AA$^{-5}$&
       b$_4$ =   135624.70 e\AA$^{-6}$&
           l$_1$ =   309.92   \AA$^{-1}$\\
    
       l$_2$ =       87.23 \AA$^{-1}$&
       l$_3$ =       32.98 \AA$^{-1}$&
       l$_4$ =        13.80 \AA$^{-1}$ \\
\end{tabular}
\label {table:slater}
\end{table}

\begin{table}[htp]
\caption{ Comparison of defect formation and migration energies (all in eV), between
          our values and best values as per previous potentials,\cite{morelon_pot} compared
          with \textit{ab initio} values from this work and with experimental values.\cite{fink,govers1} }
\centering
\begin{tabular}{c| c| c |c}
\hline\hline

\hline
       &
      Exptl.(E)/ \textit{ab initio}(AI) &
       This work &
       Previous Potentials \\
\hline
        Oxygen Frenkel Pair Formation Energy &
       3.5 +/- 0.5(E), 3.9 (AI) &
   3.3 &
       3.17\\
    \hline
               Uranium Frenkel Pair Formation Energy &
       9.5 -12(E),10.1 (AI) &
   15.5 &
       12.6\\
       
       \hline
               Schottky Trio Formation Energy &
     6.5+/- 0.5(E),7.4 (AI) &
   7.1 &
       6.68\\
\hline      
               Oxygen Interstitial Migration Energy &
     0.9-1.3(E),1.5(AI) &
   1.4 &
       0.65\\ 
\hline
        Oxygen Vacancy Migration Energy &
     0.5(E),0.9(AI) &
   0.5 &
       0.33\\ 
\hline      
               Uranium Interstitial Migration Energy &
     2.0(E),1.3(AI) &
   2.3 &
       5.0\\ 
\hline
        Uranium Vacancy Migration Energy &
     2.5(E),2.8(AI) &
   2.8 &
       4.5\\        
       
\end{tabular}
\label {table:energies}
\end{table}

\begin{table}[htp]
\caption{ Comparison of various ground state elastic properties, between
          our values and best values as per previous potentials,\cite{morelon_pot} compared
           with (extrapolated) experimental values.\cite{fink} N.B. These are \textit{predicted} and not \textit{fitted} values.}
\centering
\begin{tabular}{c | c | c | c }
\hline\hline

\hline
       &
       Exptl. &
       This work &
       Previous  potentials\\
       
       \hline

        Lattice Parameter (\AA) &
       5.46 &
       5.46 &
       5.46 \\
    \hline
              Bulk Modulus (GPa) &
       207 &
      210 &
      125 \\
      \hline
          Elastic Constant $C_{11}$ (GPa) &
       389.3 &
      401.8 &
      216.9 \\
      \hline
           Elastic Constant $C_{12}$ (GPa) &
       118.7 &
      114.1 &
      79.1 \\
      \hline
           Elastic Constant $C_{44}$ (GPa) &
       59.7 &
      107.8 &
      78.5 \\
\end{tabular}
\label {table:elastic}
\end{table}

\begin{thebibliography}{24}
\expandafter\ifx\csname natexlab\endcsname\relax\def\natexlab#1{#1}\fi
\expandafter\ifx\csname bibnamefont\endcsname\relax
  \def\bibnamefont#1{#1}\fi
\expandafter\ifx\csname bibfnamefont\endcsname\relax
  \def\bibfnamefont#1{#1}\fi
\expandafter\ifx\csname citenamefont\endcsname\relax
  \def\citenamefont#1{#1}\fi
\expandafter\ifx\csname url\endcsname\relax
  \def\url#1{\texttt{#1}}\fi
\expandafter\ifx\csname urlprefix\endcsname\relax\def\urlprefix{URL }\fi
\providecommand{\bibinfo}[2]{#2}
\providecommand{\eprint}[2][]{\url{#2}}

\bibitem[{\citenamefont{Van~Brutzel et~al.}(2008)\citenamefont{Van~Brutzel,
  Chartier, and Crocombette}}]{Brutzel_rad_damage}
\bibinfo{author}{\bibfnamefont{L.}~\bibnamefont{Van~Brutzel}},
  \bibinfo{author}{\bibfnamefont{A.}~\bibnamefont{Chartier}}, \bibnamefont{and}
  \bibinfo{author}{\bibfnamefont{J.-P.} \bibnamefont{Crocombette}},
  \bibinfo{journal}{Phys. Rev. B} \textbf{\bibinfo{volume}{78}},
  \bibinfo{pages}{024111} (\bibinfo{year}{2008}).

\bibitem[{\citenamefont{Ziegler et~al.}(Pergamon, New York,
  1985)\citenamefont{Ziegler, Biersack, and Littmark}}]{zbl}
\bibinfo{author}{\bibfnamefont{J.~F.} \bibnamefont{Ziegler}},
  \bibinfo{author}{\bibfnamefont{J.~P.} \bibnamefont{Biersack}},
  \bibnamefont{and} \bibinfo{author}{\bibfnamefont{U.}~\bibnamefont{Littmark}},
  \emph{\bibinfo{title}{The stopping and range of ions in matter}}
  (\bibinfo{year}{Pergamon, New York, 1985}).

\bibitem[{\citenamefont{Brandt et~al.}(1982)\citenamefont{Brandt and Kitagawa}}]{brandt}
\bibinfo{author}{\bibfnamefont{W.}~\bibnamefont{Brandt}}, \bibnamefont{and}
  \bibinfo{author}{\bibfnamefont{J.-P.} \bibnamefont{Kitagawa}},
  \bibinfo{journal}{Phys. Rev. B} \textbf{\bibinfo{volume}{25}},
  \bibinfo{pages}{5631} (\bibinfo{year}{1982}).
    
    \bibitem[{\citenamefont{Cai et~al.}(1996)\citenamefont{Cai, Gr{\o}nbech-Jensen, Snell, and Beardmore}}]{cai}
    \bibinfo{author}{\bibfnamefont{D.}~\bibnamefont{Cai}},
\bibinfo{author}{\bibfnamefont{N.}~\bibnamefont{Gr{\o}nbech-Jensen}},
  \bibinfo{author}{\bibfnamefont{C.~M.}~\bibnamefont{Snell}}, \bibnamefont{and}
  \bibinfo{author}{\bibfnamefont{K.~M.} \bibnamefont{Beardmore}},
  \bibinfo{journal}{Phys. Rev. B} \textbf{\bibinfo{volume}{54}},
  \bibinfo{pages}{17147} (\bibinfo{year}{1996}).
    
    
    \bibitem[{\citenamefont{Barberan et~al.}(1986)\citenamefont{Barberan and Echenique}}]{barberan}
\bibinfo{author}{\bibfnamefont{N.}~\bibnamefont{Barberan}}, \bibnamefont{and}
  \bibinfo{author}{\bibfnamefont{P.~M.} \bibnamefont{Echenique}},
  \bibinfo{journal}{Jour. Phys. B: Atomic and Molecular Physics} \textbf{\bibinfo{volume}{19}},
  \bibinfo{pages}{L81} (\bibinfo{year}{1986}).


    \bibitem[{\citenamefont{Wilson et~al.}(1980)\citenamefont{Wilson and Deline}}]{wilson}
\bibinfo{author}{\bibfnamefont{R.~G.}~\bibnamefont{Wilson}}, \bibnamefont{and}
  \bibinfo{author}{\bibfnamefont{V.~R.} \bibnamefont{Deline}},
  \bibinfo{journal}{Appl. Phys. Lett.} \textbf{\bibinfo{volume}{37}},
  \bibinfo{pages}{793} (\bibinfo{year}{1980}).
  
  
\bibitem[{\citenamefont{Nastasi et~al.}(Cambridge University Press, New York,
  1996)\citenamefont{Nastasi, Mayer, and Hirvonen}}]{nastasi}
\bibinfo{author}{\bibfnamefont{M.} \bibnamefont{Nastasi}},
  \bibinfo{author}{\bibfnamefont{J.~W.} \bibnamefont{Mayer}},
  \bibnamefont{and} \bibinfo{author}{\bibfnamefont{J.~K.}~\bibnamefont{Hirvonen}},
  \emph{\bibinfo{title}{Ion-solid interactions: Fundamentals and applications}}
  (\bibinfo{year}{Cambridge University Press, New York, 1996}).
    
\bibitem[{\citenamefont{Geng et~al.}(2008)\citenamefont{Geng, Chen, Kaneta,
  Iwasawa, Ohnuma, and Kinoshita}}]{geng2008}
\bibinfo{author}{\bibfnamefont{H.~Y.} \bibnamefont{Geng}},
  \bibinfo{author}{\bibfnamefont{Y.}~\bibnamefont{Chen}},
  \bibinfo{author}{\bibfnamefont{Y.}~\bibnamefont{Kaneta}},
  \bibinfo{author}{\bibfnamefont{M.}~\bibnamefont{Iwasawa}},
  \bibinfo{author}{\bibfnamefont{T.}~\bibnamefont{Ohnuma}}, \bibnamefont{and}
  \bibinfo{author}{\bibfnamefont{M.}~\bibnamefont{Kinoshita}},
  \bibinfo{journal}{Phys. Rev. B} \textbf{\bibinfo{volume}{77}},
  \bibinfo{pages}{104120} (\bibinfo{year}{2008}).

\bibitem[{\citenamefont{Van~Brutzel et~al.}(2003)\citenamefont{Van~Brutzel,
  Delaye, Ghaleb, and Rarivomanantsoa}}]{brutzel_md}
\bibinfo{author}{\bibfnamefont{L.}~\bibnamefont{Van~Brutzel}},
  \bibinfo{author}{\bibfnamefont{J.-M.} \bibnamefont{Delaye}},
  \bibinfo{author}{\bibfnamefont{D.}~\bibnamefont{Ghaleb}}, \bibnamefont{and}
  \bibinfo{author}{\bibfnamefont{M.}~\bibnamefont{Rarivomanantsoa}},
  \bibinfo{journal}{Phil. Mag.} \textbf{\bibinfo{volume}{83}},
  \bibinfo{pages}{4083} (\bibinfo{year}{2003}).


\bibitem[{\citenamefont{}(2007)}]{govers1}
\bibinfo{author}{\bibfnamefont{K.} \bibnamefont{Govers}},
\bibinfo{author}{\bibfnamefont{S.} \bibnamefont{Lemehov}},
\bibinfo{author}{\bibfnamefont{M.} \bibnamefont{Hou}},
\bibinfo{author}{\bibfnamefont{M.} \bibnamefont{Verwerft}},
 \bibinfo{journal}{J.
  Nucl. Mater.} \textbf{\bibinfo{volume}{366}}, \bibinfo{pages}{161}
  (\bibinfo{year}{2007}).
  
  \bibitem[{\citenamefont{}(2008)}]{govers2}
\bibinfo{author}{\bibfnamefont{K.} \bibnamefont{Govers}},
\bibinfo{author}{\bibfnamefont{S.} \bibnamefont{Lemehov}},
\bibinfo{author}{\bibfnamefont{M.} \bibnamefont{Hou}},
\bibinfo{author}{\bibfnamefont{M.} \bibnamefont{Verwerft}},
 \bibinfo{journal}{J.
  Nucl. Mater.} \textbf{\bibinfo{volume}{376}}, \bibinfo{pages}{66}
  (\bibinfo{year}{2008}).
  
\bibitem[{\citenamefont{Slater}(1932)}]{slater}
\bibinfo{author}{\bibfnamefont{J.~C.} \bibnamefont{Slater}},
  \bibinfo{journal}{Phys. Rev.} \textbf{\bibinfo{volume}{42}},
  \bibinfo{pages}{33} (\bibinfo{year}{1932}).

\bibitem[{\citenamefont{Laskowski et~al.}(2004)\citenamefont{Laskowski, Madsen,
  Blaha, and Schwarz}}]{laskowski}
\bibinfo{author}{\bibfnamefont{R.}~\bibnamefont{Laskowski}},
  \bibinfo{author}{\bibfnamefont{G.~K.~H.} \bibnamefont{Madsen}},
  \bibinfo{author}{\bibfnamefont{P.}~\bibnamefont{Blaha}}, \bibnamefont{and}
  \bibinfo{author}{\bibfnamefont{K.}~\bibnamefont{Schwarz}},
  \bibinfo{journal}{Phys. Rev. B} \textbf{\bibinfo{volume}{69}},
  \bibinfo{pages}{140408} (\bibinfo{year}{2004}).

\bibitem[{\citenamefont{Geng et~al.}(2007)\citenamefont{Geng, Chen, Kaneta, and
  Kinoshita}}]{geng2007}
\bibinfo{author}{\bibfnamefont{H.~Y.} \bibnamefont{Geng}},
  \bibinfo{author}{\bibfnamefont{Y.}~\bibnamefont{Chen}},
  \bibinfo{author}{\bibfnamefont{Y.}~\bibnamefont{Kaneta}}, \bibnamefont{and}
  \bibinfo{author}{\bibfnamefont{M.}~\bibnamefont{Kinoshita}},
  \bibinfo{journal}{Phys. Rev. B} \textbf{\bibinfo{volume}{75}},
  \bibinfo{pages}{054111} (\bibinfo{year}{2007}).

\bibitem[{\citenamefont{Ikushima et~al.}(2001)\citenamefont{Ikushima, Tsutsui,
  Haga, Yasuoka, Walstedt, Masaki, Nakamura, Nasu, and \ifmmode~\bar{O}\else
  \={O}\fi{}nuki}}]{ikushima}
\bibinfo{author}{\bibfnamefont{K.}~\bibnamefont{Ikushima}},
  \bibinfo{author}{\bibfnamefont{S.}~\bibnamefont{Tsutsui}},
  \bibinfo{author}{\bibfnamefont{Y.}~\bibnamefont{Haga}},
  \bibinfo{author}{\bibfnamefont{H.}~\bibnamefont{Yasuoka}},
  \bibinfo{author}{\bibfnamefont{R.~E.} \bibnamefont{Walstedt}},
  \bibinfo{author}{\bibfnamefont{N.~M.} \bibnamefont{Masaki}},
  \bibinfo{author}{\bibfnamefont{A.}~\bibnamefont{Nakamura}},
  \bibinfo{author}{\bibfnamefont{S.}~\bibnamefont{Nasu}}, \bibnamefont{and}
  \bibinfo{author}{\bibfnamefont{Y.}~\bibnamefont{\ifmmode~\bar{O}\else
  \={O}\fi{}nuki}}, \bibinfo{journal}{Phys. Rev. B}
  \textbf{\bibinfo{volume}{63}}, \bibinfo{pages}{104404}
  (\bibinfo{year}{2001}).

\bibitem[{\citenamefont{Kresse and Furthm\"uller}(1996)}]{VASP}
\bibinfo{author}{\bibfnamefont{G.}~\bibnamefont{Kresse}} \bibnamefont{and}
  \bibinfo{author}{\bibfnamefont{J.}~\bibnamefont{Furthm\"uller}},
  \bibinfo{journal}{Phys. Rev. B} \textbf{\bibinfo{volume}{54}},
  \bibinfo{pages}{11169} (\bibinfo{year}{1996}).

\bibitem[{\citenamefont{Anisimov et~al.}(1991)\citenamefont{Anisimov, Zaanen,
  and Andersen}}]{lda+u}
\bibinfo{author}{\bibfnamefont{V.~I.} \bibnamefont{Anisimov}},
  \bibinfo{author}{\bibfnamefont{J.}~\bibnamefont{Zaanen}}, \bibnamefont{and}
  \bibinfo{author}{\bibfnamefont{O.~K.} \bibnamefont{Andersen}},
  \bibinfo{journal}{Phys. Rev. B} \textbf{\bibinfo{volume}{44}},
  \bibinfo{pages}{943} (\bibinfo{year}{1991}).

\bibitem[{\citenamefont{Dudarev et~al.}(1998)\citenamefont{Dudarev, Botton,
  Savrasov, Humphreys, and Sutton}}]{dudarev1}
\bibinfo{author}{\bibfnamefont{S.~L.} \bibnamefont{Dudarev}},
  \bibinfo{author}{\bibfnamefont{G.~A.} \bibnamefont{Botton}},
  \bibinfo{author}{\bibfnamefont{S.~Y.} \bibnamefont{Savrasov}},
  \bibinfo{author}{\bibfnamefont{C.~J.} \bibnamefont{Humphreys}},
  \bibnamefont{and} \bibinfo{author}{\bibfnamefont{A.~P.}
  \bibnamefont{Sutton}}, \bibinfo{journal}{Phys. Rev. B}
  \textbf{\bibinfo{volume}{57}}, \bibinfo{pages}{1505} (\bibinfo{year}{1998}).

\bibitem[{\citenamefont{Dudarev et~al.}(1999)\citenamefont{Dudarev, Botton,
  Savrasov, Szotek, Temmerman, and Sutton}}]{dudarev2}
\bibinfo{author}{\bibfnamefont{S.~L.} \bibnamefont{Dudarev}},
  \bibinfo{author}{\bibfnamefont{G.~A.} \bibnamefont{Botton}},
  \bibinfo{author}{\bibfnamefont{S.~Y.} \bibnamefont{Savrasov}},
  \bibinfo{author}{\bibfnamefont{Z.}~\bibnamefont{Szotek}},
  \bibinfo{author}{\bibfnamefont{W.~M.} \bibnamefont{Temmerman}},
  \bibnamefont{and} \bibinfo{author}{\bibfnamefont{A.~P.}
  \bibnamefont{Sutton}}, \bibinfo{journal}{Phys. Stat. Sol. A}
  \textbf{\bibinfo{volume}{166}}, \bibinfo{pages}{429} (\bibinfo{year}{1999}).

\bibitem[{\citenamefont{Jonsson et~al.}(World Scientific, Singapore,
  1984)\citenamefont{Jonsson, Mills, and Jacobsen}}]{neb}
\bibinfo{author}{\bibfnamefont{H.}~\bibnamefont{Jonsson}},
  \bibinfo{author}{\bibfnamefont{G.}~\bibnamefont{Mills}}, \bibnamefont{and}
  \bibinfo{author}{\bibfnamefont{K.~W.} \bibnamefont{Jacobsen}},
  \emph{\bibinfo{title}{in classical and quantum dynamics in condensed phase
  simulations}}, p. \bibinfo{pages}{385} (\bibinfo{year}{World Scientific,
  Singapore, 1984}).

\bibitem[{\citenamefont{Henkelman et~al.}(2000)\citenamefont{Henkelman,
  Uberuaga, and Jonsson}}]{climbingimage}
\bibinfo{author}{\bibfnamefont{G.}~\bibnamefont{Henkelman}},
  \bibinfo{author}{\bibfnamefont{B.~P.} \bibnamefont{Uberuaga}},
  \bibnamefont{and} \bibinfo{author}{\bibfnamefont{H.}~\bibnamefont{Jonsson}},
  \bibinfo{journal}{J. Chem. Phys.} \textbf{\bibinfo{volume}{113}},
  \bibinfo{pages}{9901} (\bibinfo{year}{2000}).

\bibitem[{\citenamefont{Matzke}(1987)}]{Matzke}
\bibinfo{author}{\bibfnamefont{H.} \bibnamefont{Matzke}}, \bibinfo{journal}{J.
  Chem, Soc., Faraday Trans.} \textbf{\bibinfo{volume}{83}},
  \bibinfo{pages}{1121} (\bibinfo{year}{1987}).
  
\bibitem[{\citenamefont{Morelon et~al.}(2003)\citenamefont{Morelon, Ghaleb,
  Delaye, and Van~Brutzel}}]{morelon_pot}
\bibinfo{author}{\bibfnamefont{N.~D.} \bibnamefont{Morelon}},
  \bibinfo{author}{\bibfnamefont{D.}~\bibnamefont{Ghaleb}},
  \bibinfo{author}{\bibfnamefont{J.-M.} \bibnamefont{Delaye}},
  \bibnamefont{and}
  \bibinfo{author}{\bibfnamefont{L.}~\bibnamefont{Van~Brutzel}},
  \bibinfo{journal}{Phil. Mag.} \textbf{\bibinfo{volume}{83}},
  \bibinfo{pages}{1533} (\bibinfo{year}{2003}).

\bibitem[{\citenamefont{Nelder and Mead}(1965)}]{amoeba}
\bibinfo{author}{\bibfnamefont{J.~A.} \bibnamefont{Nelder}} \bibnamefont{and}
  \bibinfo{author}{\bibfnamefont{R.}~\bibnamefont{Mead}}, \bibinfo{journal}{The
  Comp. Jour.} \textbf{\bibinfo{volume}{7}}, \bibinfo{pages}{308}
  (\bibinfo{year}{1965}).

\bibitem[{\citenamefont{Gale}(1997)}]{GULP}
\bibinfo{author}{\bibfnamefont{J.~D.} \bibnamefont{Gale}}, \bibinfo{journal}{J.
  Chem, Soc., Faraday Trans.} \textbf{\bibinfo{volume}{93}},
  \bibinfo{pages}{629} (\bibinfo{year}{1997}).


\bibitem[{\citenamefont{Fink}(2000)}]{fink}
\bibinfo{author}{\bibfnamefont{J.~K.} \bibnamefont{Fink}}, \bibinfo{journal}{J.
  Nucl. Mater.} \textbf{\bibinfo{volume}{279}}, \bibinfo{pages}{1}
  (\bibinfo{year}{2000}).

\bibitem[{\citenamefont{Karakasidis and Lindan}(1994)}]{old_pot_1}
\bibinfo{author}{\bibfnamefont{T.}~\bibnamefont{Karakasidis}} \bibnamefont{and}
  \bibinfo{author}{\bibfnamefont{P.~J.~D.} \bibnamefont{Lindan}},
  \bibinfo{journal}{J. Phys: Condens. Matter} \textbf{\bibinfo{volume}{6}},
  \bibinfo{pages}{2965} (\bibinfo{year}{1994}).

\bibitem[{\citenamefont{Sindzingre and Gillan}(1988)}]{old_pot_2}
\bibinfo{author}{\bibfnamefont{P.}~\bibnamefont{Sindzingre}} \bibnamefont{and}
  \bibinfo{author}{\bibfnamefont{M.~J.} \bibnamefont{Gillan}},
  \bibinfo{journal}{J. Phys. C: Solid State Phys.}
  \textbf{\bibinfo{volume}{21}}, \bibinfo{pages}{4017} (\bibinfo{year}{1988}).

\end{thebibliography}
\end{document}